\documentclass[12pt,a4paper]{article}
\usepackage{graphicx}
\usepackage{amsmath}
\usepackage{amssymb}
\usepackage{epsfig}
\usepackage{times}
\begin{document}

\title{Einstein's `Z\"urich  Notebook' and his Journey to General Relativity}
\author{Norbert Straumann \\
Institute for Theoretical Physics University of Z\"urich,\\
Winterthurerstrasse 190, CH--8057 Z\"urich, Switzerland}
%\date{}
\maketitle
\begin{abstract}
On the basis of his `Z\"urich  Notebook' I shall describe a particularly fruitful phase in Einstein's struggle on the way to general relativity. These research notes are an extremely illuminating source for understanding Einstein's main physical arguments and conceptual difficulties that delayed his discovery of general relativity by about three years. Together with the `Ent\-wurf' theory in collaboration with Marcel Grossmann, these notes also show that the final theory was missed late in 1912 within a hair's breadth. The Einstein-Grossmann theory, published almost exactly hundred years ago, contains, however, virtually all essential elements of Einstein's definite gravi\-tation theory. This should become obvious by our streamlined presentation of the final phase in November 1915 that culminated with the definite field equations.
\end{abstract}

\section{Introduction}

\label{intro}
Einstein's path to general relativity (GR) meandered steeply, encountered confusing forks, and also included a big U-turn.
Einstein's own words to describe the ambivalent feelings of the searching mind are unforgettable \cite{GS}:

\begin{quote}
In the light of knowledge attained, the happy achievement seems almost a matter of course, and any intelligent student can grasp it without too much trouble. But the years of anxious searching in the dark, with their intense longing, their alternation of confidence and exhaustion and the final emergence into light -- only those who have experienced it can understand it.
\end{quote}

This is not the place to give an account of the complex history that led from special relativity (SR) to general relativity in the course of about eight years. What I will do in this article is to discuss in some detail Einstein's remarkable progress beginning in August 1912, after his second return to Z\"urich, until Spring 1913. Before I come to this, I should presumably indicate what he had already achieved before this period.

In 1907, while writing a review article on SR, Einstein speculated -- attempting to understand the empirical equality of inertial and gravitational mass -- on the possibility of extending the principle of relativity to accelerated motion, and added an important section on gravitation in his review \cite{EqP}.\footnote{References to papers that have appeared in the \textit{Collected Papers of Albert Einstein} (CPAE) \cite{Ein0} are always cited by volume and document of CPAE.}  With this ``basic idea'', which he referred to as \textit{principle of equivalence}, he went beyond the framework of SR. Indeed, he did not seriously consider the possibility of
a special-relativistic theory of gravity until presented with such a theory by Gunnar Nordstr\"om. Except for his attempted rebuttals of Nordstr\"om's
theories, no notes appear to be extant to document his own early attempts in this direction. But later recollections
by Einstein make it quite easy to more or less guess the essential steps (see \cite{GS}). His (special formulation) of the equivalence principle -- ``the most fortunate thought of my life'' --  became the guiding thread in his search for a relativistic theory of gravitation.

Until 1911 Einstein worked apparently mainly on the quantum puzzles and did not publish anything about gravitation, but continued to think about the problem.  In \cite{Ein3} he writes: ``Between 1909-1912 while I had to teach theoretical physics at the Z\"urich and Prague Universities I pondered ceaselessly on the problem''. When Einstein realized in 1911 that gravitational light deflection should be experimentally observable \cite{EqP2}, he took up the problem of gravitation again and began to ``work like a horse'' in developing a coherent theory of the static gravitational fields.  Since he had found that the velocity of light depends on the gravitational potential, he concluded that the speed of light plays the role of the gravitational potential, and proposed a non-linear field equation, in which the gravitational energy density itself acts as a source of the gravitational potential. Therefore, the field equation implied that the principle of equivalence is valid only for infinitely small spatial regions. In the second of his Prague papers on ``gravito-statics'' \cite{Ein4a} he also showed how the equations of electrodynamics and thermodynamics are modified in the presence of a static gravitational field. At this point he began to investigate the dynamical gravitational field.

\section{Einstein's Z\"urich  Notebook}
\label{sec:1}
%and \cite{RefJ}

I come now to a detailed discussion of Einstein's Z\"urich  notebook \cite{Ein1}. It is really fascinating to study these research notes, because one can see Einstein at work, and theoretical physics at its best: A delicate interplay between physical reasoning, based on an intuitive estimate of the most relevant empirical facts, and -- equally important -- mathematical structural aspects and requirements.

Soon after he returned to Z\"urich  in August 1912, Einstein encountered (apparent) tension between certain physical requirements and a satisfactory mathematical formulation, based on the work of Riemann, Christoffel, Ricci, and Levi-Civita on differential covariants. We shall see that Einstein, in close collaboration with the mathematician Marcel Grossmann,  already late in 1912 came very close to his final theory, but physical and conceptual arguments convinced him for a long time that -- with ``heavy heart'' -- he had to abandon the general covariance of the gravitational field equations. In a letter to Lorentz \cite{Ein2} he called this the ``ugly dark spot'' of the theory. With this decision, based on erroneous judgement, Einstein lost almost three years until physics and mathematics came into harmony in his beautiful general theory of relativity.

Historians of science have, of course, studied the Z\"urich  Notebook extensively (see especially \cite{Nor1}, \cite{Renn}). What I will try to do is to present the main steps of Einstein's research during his time at the ETH  in a way that is hopefully appealing to active physicists with only a peripheral interest in the history of their field. In my exposition I will always use modern notation and assume a working knowledge of GR.

\subsection{Starting point in August 1912}
\label{sec:1.1}

When Einstein arrived in Z\"urich  in early August, he was convinced that a metric field of spacetime, generalizing the Minkowski metric to a pseudo-Riemannian dynamical metric, was the right relativistic generalization of Newton's potential. The main question was to find field equations for this field. But how to achieve this was in the dark and he looked for mathematical help. Fortunately, Marcel Grossmann, his old friend since his student time, was now also professor at the ETH and Einstein succeeded in gaining him as a collaborator in his search to the equations. In a 1955 reminiscence, shortly before his death, Einstein wrote \cite{Ein3}:

\begin{quote}
I was made aware of these [works by Ricci and Levi-Civita] by my friend Grossmann in Z\"urich, when I put the problem to investigate generally covariant tensors, whose components depend only on the derivatives of the coefficients of the quadratic fundamental invariant.

He at once caught fire, although as a mathematician he had a somewhat sceptical stance towards physics. (...) He went through the literature and soon discovered that the indicated mathematical problem had already been solved, in particular by Riemann, Ricci and Levi-Civita. This entire development was connected to the Gaussian theory of curved surfaces, in which for the first time systematic use was made of generalized coordinates. 
\end{quote} 

Louis Kollros, another student friend of Einstein, who was also mathematics professor at the ETH during this time, remembered also in 1955 \cite{Koll}: 

\begin{quote}
[Einstein] spoke to Grossmann about his troubles and said one day: ``Grossmann, you must help me, otherwise I'll go crazy ! ''.
\end{quote}

Already on the first page of the Z\"urich  Notebook Einstein derives the transformation law for the coefficients $g_{\mu\nu}$ of the metric under general (smooth) coordinate transformations. On the same page he also writes down the non-linear field equation he had obtained in Prague for static fields, assuming that the spatial metric is\textit{ flat}. (This assumption will later play a crucial, but unfortunate role.) In the same paper he had derived the equation of motion for a point particle from a variational principle, which is now generalized in an natural manner to 

\begin{equation}
\delta\int ds=0 , \qquad ds^2=g_{\mu\nu}dx^\mu dx^\nu.
\label{Eq:1}
\end{equation}

\subsection{Requirements to be satisfied by the future theory}
\label{sec:1.2}

The following mixture of physical and mathematical properties of a relativistic theory of gravitation are among Einstein's main guiding principles:

$\bullet$ The theory reduces to the Newtonian limit for weak fields and slowly moving matter.

$\bullet$ Conservation laws for energy and momentum must hold.

$\bullet$ The equivalence principle must be embodied.

$\bullet$ The theory respects a generalized principle of relativity to accelerating frames, taking into account that gravitation and inertia are described by one and the same field $g_{\mu\nu}$. Einstein expressed this by the requirement of general covariance of the basic equations (to become a much debated subject).

\subsection{Energy-momentum conservation for dust}
\label{sec:1.3}

Early in the Z\"urich  Notebook Einstein writes the geodesic equation of motion for a point particle in the form
\begin{equation}
\frac{d}{d\tau}\Bigl(g_{\mu\nu} \frac{dx^\nu}{d\tau}\Bigr) -\frac{1}{2}\partial_\mu g_{\alpha\beta}\frac{dx^\alpha}{d\tau}\frac{dx^\beta}{d\tau}=0 .
\label{Eq:2}
\end{equation}
Considering an incoherent dust distribution as an ensemble of particles, he guesses that the energy-momentum conservation law of special relativity, $\partial_\nu T^{\mu\nu}=f^\mu$, with the energy-stress tensor $T^{\mu\nu}=\rho_0 u^\mu u^\nu,\,(\rho_0$ = rest mass matter density, $u^\mu$ = four-velocity field) and an external force density $f^{\mu}$, should be replaced by 
\begin{equation}
\frac{1}{\sqrt{-g}}\partial_\nu(\sqrt{-g}g_{\mu\lambda}T^{\lambda\nu})-\frac{1}{2}\partial_{\mu}g_{\alpha\beta}T^{\alpha\beta}=0
\label{Eq:3}
\end{equation}
($g:=\det (g_{\mu\nu})$).
The details of Einstein's considerations are described in his Part I, Sect. 4 of the ``Entwurf'' paper by Einstein and Grossmann \cite{EG}. We know, of course, that this is just an explicit form of the equation $\nabla_\nu T_{\mu}{}^{\nu}=0$, and this is also stated by Grossmann in his Part II of \cite{EG}. In his notes Einstein checks the general invariance of (\ref{Eq:3}). (Don't forget, Einstein was at this point still a beginner in the use of the absolute differential calculus.) He asks whether the left hand side of the equation, generated from a symmetric contravariant tensor field, is always a vector field. (I am simplifying things a bit.) If so, this would have to be the case for the tensor field $g^{\mu\nu}$. With a simple calculation he shows that for this example he gets zero (metric condition); this is at least a consistency test. 

What Einstein presents is not a derivation, but a natural guess.\footnote{A modern author might translate Einstein's reasoning using kinetic theory. If the distribution function satisfies the collisionless Boltzmann equation, Einstein's formula  (\ref{Eq:3}) follows. This is, for instance, shown in Chap. 7 of \cite{NS}. The simple form 
$T^{\mu\nu}=\rho_0 u^\mu u^\nu$ requires, however, additional assumptions: $T^{\mu\nu}$ is given by a second moment, while the right hand side is a product of first moments. For a different extensive discussion of Einstein's guess, see \cite{Nor2}, especially Appendices B and C.} We would now add that (\ref{Eq:3}) implies that the integral curves of the four-velocity field $u^{\mu}$ are geodesics for incoherent (pressureless) dust.

\section{In search of the gravitational field equations}
\label{sec:2}

Soon, Einstein begins to look for candidate field equations. The pages before 27 of the Z\"urich  Notebook show that he was not yet acquainted with the absolute calculus of Ricci and Levi-Civita. On p. 26 he considers for the case $-g=1$ the equation
\begin{equation}
 g^{\alpha\beta}\partial_{\alpha} \partial_{\beta}g^{\mu\nu}=\kappa T^{\mu\nu},
 \label{4.0}
 \end{equation}
and substitutes the left hand side for $T^{\mu\nu}$ in (\ref{Eq:3}), but that produces third derivatives and leads to nowhere.

\subsection{Einstein studies the Ricci tensor as a candidate}
\label{sec:2.1}

On p. 27, referring to Grossmann, Einstein writes down the expression for the fully covariant Riemann curvature tensor $R_{\alpha\beta\gamma\delta}$. Next, he forms by contraction the Ricci tensor $R_{\mu\nu}$. The resulting terms involving second derivatives consist, beside $ g^{\alpha\beta}\partial_{\alpha} \partial_{\beta}g_{\mu\nu}$, of three additional terms. Einstein writes below their sum: ``should vanish'' [``sollte verschwinden'']. The reason is that he was looking for field equations of the following general form:
\begin{equation}
\Gamma^{\mu\nu}[g]=\kappa T^{\mu\nu},
\label{Eq:4}
\end{equation}
with 
\begin{equation}
\Gamma^{\mu\nu}[g]=\partial_{\alpha}(g^{\alpha\beta}\partial_{\beta}g^{\mu\nu}) + \mbox{terms that vanish in linear approximation}. 
\label{Eq:5}
\end{equation}
On the next page he considers the curvature scalar $R$, and writes it out explicitly in terms of the metric. This calculation goes on for several pages. The final expression for $R$ assumes that the coordinate system is unimodular ($-g=1$). Then Einstein starts again in writing out $R^{\mu\nu}$ explicitly in terms of $g_{\mu\nu}$ (inserting the expressions for the Christoffel symbols), and runs for the non-linear terms into a mess, commented as ``too complicated'' [``zu umst\"andlich''].  On p. 37 he begins with a new attempt,  but simplifies this time the result for $R_{\mu\nu}$ in coordinates that satisfy the \textit{harmonic condition}
\begin{equation}
\square x^\alpha=0,\quad \square:=\frac{1}{\sqrt{-g}}\partial_\mu(\sqrt{-g}g^{\mu\nu}\partial_\nu)
\label{Eq:6}
\end{equation}
or $\Gamma^\alpha=0$, where
\begin{equation}
\Gamma^\alpha:=g^{\mu\nu}\Gamma^{\alpha}{}_{\mu\nu}=-\partial_{\mu}g^{\mu\alpha}-\frac{1}{2}g^{\alpha\beta}g^{\mu\nu}\partial_{\beta}g_{\mu\nu}.
\label{Eq:7}
\end{equation}
Einstein notes that with this coordinate choice the only  term with second derivatives is now $-(1/2)g^{\alpha\beta}\partial_\alpha\partial_\beta g_{\mu\nu}$, and therefore the result is of the desired form (\ref{Eq:4}, \ref{Eq:5}): In harmonic coordinates\footnote{In general coordinates the Ricci tensor is given by \[ R_{\mu\nu}=\,^{(h)\!}R_{\mu\nu}+\frac{1}{2}(g_{\alpha\mu} \partial_{\nu}\Gamma^{\alpha}+g_{\alpha\nu} \partial_{\mu}\Gamma^{\alpha}).\]}:
\begin{equation}
\,^{(h)\!}R_{\mu\nu}=-\frac{1}{2}g^{\alpha\beta}\partial_\alpha\partial_\beta g_{\mu\nu}+H_{\mu\nu}(g,\partial g),
\label{Eq:8 }
\end{equation}
where $H_{\mu\nu}(g,\partial g)$ is a rational expression of $g_{\mu\nu}$ and $\partial_{\alpha}g_{\mu\nu}$ (with denominator $g$) that vanishes in the linear approximation. This is, of course, a familiar result for us which plays an important role in GR (for instance, in studying the Cauchy problem).

This seems to look good, and Einstein begins to analyse the linear weak field approximation of the field equations\footnote{Never before had Einstein used in his work such advanced and complex mathematics. This is expressed in a letter to Arnold Sommerfeld on 29 October 1912 (CPAE, Vol. 5, Doc. 421): ``But one thing is certain: never before in my life have I toiled any where near as much, and I have gained enormous respect for mathematics, whose more subtle parts I considered until now, in my ignorance, as pure luxury. Compared with this problem, the original theory of relativity is child's play.''}
\begin{equation}
R_{\mu\nu}=\kappa T_{\mu\nu}.
\label{Eq:9}
\end{equation}
(We know, of course, that he has to run into problems, because of the contracted Bianchi identity.)

\subsection{The weak field approximation}
\label{sec:2.2}

The linearized harmonic coordinate condition becomes for $h_{\mu\nu}:=g_{\mu\nu}-\eta_{\mu\nu}$ ($\eta_{\mu\nu}$: Minkowski metric)
\begin{equation}
\partial_{\mu}(h^{\mu\alpha}-\frac{1}{2}\eta^{\mu\alpha}h)=0
\label{Eq:10}
\end{equation}
($h:=h^{\mu}{}_{\mu}$, indices are now raised and lowered by means of the Minkowski metric). This is nowadays usually called the \textit{Hilbert condition}, but Einstein imposed it already in 1912. The field equations become
\begin{equation}
\square h_{\mu\nu}=-2\kappa T_{\mu\nu}.
\label{Eq:11}
\end{equation}
Einstein takes for $T_{\mu\nu}$ his earlier expression for dust.

But now he runs into a \textbf{serious problem}:

From $\partial^{\nu}T_{\mu\nu}=0$ in the weak field limit, it follows that $\square(\partial^{\nu}h_{\mu\nu})=0$, hence the harmonic coordinate condition requires $\square (\partial_{\nu}h)=0$, and therefore the trace of the the field equations implies $\square h=-2\kappa T= const.,\, T:=T^{\mu}{}_{\mu}$. For dust this requires that $T=-\rho_0 =const.$ This is, of course, unacceptable. One would not even be able to describe a star, with a smooth distribution of matter localized in a finite region of space.

It may be helpful, to point out the non-linear version of this difficulty. Equation (\ref{Eq:9}) together with $\nabla^{\nu}T_{\mu\nu}=0$ imply, using the contracted Bianchi identity $\nabla^{\nu}R_{\mu\nu}=\frac{1}{2}\partial_{\mu}R$, that $R=const.$, thus the trace of (\ref{Eq:9}) leads again to $T=const.$ Einstein discovered this, without knowing the Bianchi identity, in the fall of 1915, when he reconsidered the candidate field equations (\ref{Eq:9}).

\textit{Remark}. From his studies of static gravity in Prague, Einstein was convinced that in the (weak) static limit the metric must be of the form $(g_{\mu\nu})=\mbox{diag}(g_{00} (\mathbf{x}),1,1,1)$, thus \textit{spatially flat}. But then $\square h=const.$ would imply that $\triangle g_{00}=const.$ If the function $ g_{00}$ is bounded on $\mathbb{R}^{3}$ (in the weak field approximation the absolute value of this function should be close to 1), then $g_{00}(\mathbf{x})$ would have to be a constant.\footnote{A non-linear version of this remark may be of some interest. If the metric is assumed to be static with flat spatial sections, then we obtain in  coordinates adapted to the static Killing field for the curvature scalar 
\[R=-\frac{2}{\varphi}\triangle\varphi, \] 
with $g_{00}=:-\varphi^2$ (see \cite{Book}, Sect. 2.1). Since $R$ is constant, we obtain the equation $\triangle\varphi=\Lambda\varphi$, where the constant $\Lambda$ is equal to $-\kappa T/2$. For `normal' matter $\Lambda$ is non-negative. If $\Lambda >0 \,  (T\neq 0$) we conclude that $\varphi=0$. Since $\varphi$ must be everywhere positive, it follows that a bounded $\varphi$ has to be a constant, hence only the Minkowski metric remains. (In the non-linear case unbounded harmonic functions are, a priori, allowed.)} (Use the fact that a bounded harmonic function on $\mathbb{R}^{3}$ is constant.) 

\subsection{Einstein's modified linearized field equations}
\label{sec:2.3}

Now, something very interesting happens. Einstein avoids the first problem by modifying the field equations (\ref{Eq:11}) to
\begin{equation}
\square (h_{\mu\nu} - \frac{1}{2}\eta_{\mu\nu}h)=-2\kappa T_{\mu\nu}.
\label{Eq:12}
\end{equation}
Then the harmonic coordinate condition (\ref{Eq:10}) is compatible with $\partial_{\nu}T^{\mu\nu}=0$. Remarkably, (\ref{Eq:12}) is the linearized equation of the final theory (in harmonic coordinates). One wonders why Einstein did not try at this point the analogous substitution $R_{\mu\nu} \longrightarrow R_{\mu\nu}-\frac{1}{2}g_{\mu\nu}R$ or $T_{\mu\nu} \longrightarrow T_{\mu\nu}-\frac{1}{2}g_{\mu\nu}T$ in the full non-linear equation (\ref{Eq:9}). Before we discuss the probable reason for this, we go on with his research notes.

\subsubsection{Energy-momentum conservation for weak fields}

In linearized approximation (\ref{Eq:3})  becomes
\begin{equation}
\partial_{\nu}T_{\mu}{}^{\nu}-\frac{1}{2}\partial_{\mu}h_{\alpha\beta}T^{\alpha\beta}=0.
\label{Eq:13} 
\end{equation}
Einstein replaces in the second term $T^{\alpha\beta}$ by $(-1/2\kappa)$ times the left hand side of the modified field equations (\ref{Eq:12}). The resulting expression is proportional to the left hand side of the next equation. This is rewritten as a total divergence by performing several partial integrations:
\begin{equation}
\square(h_{\mu\nu} - \frac{1}{2}\eta_{\mu\nu}h)h^{\mu\nu}{}_{,\sigma}=-4\kappa t_{\sigma}{}^{\lambda}{}_{,\lambda},
\label{Eq:14}
\end{equation} 
 where
 \begin{equation}
 t_{\sigma}{}^{\lambda}=-\frac{1}{4\kappa}\Bigl[h_{\mu\nu}{}^{,\lambda}h^{\mu\nu}{}_{,\sigma}-\frac{1}{2}\delta^{\lambda}_{\sigma}h_{\mu\nu,\rho}h^{\mu\nu,\rho}-
 \frac{1}{2}(h^{,\lambda}h_{,\sigma}-\frac{1}{2}\delta^{\lambda}_{\sigma}h_{,\rho}h^{,\rho})\Bigr].
 \label{Eq:15}
 \end{equation}
With this identity Einstein obtains the \textit{conservation law}
\begin{equation}
\partial_{\nu}(T_{\mu}{}^{\nu}+t_{\mu}{}^{\nu})=0.
\label{Eq:16}
\end{equation}
Remember that this holds only in harmonic coordinates. (We would now add that $t_{\mu}{}^{\nu}$ is not gauge invariant, i.e., not invariant under the substitution $h_{\mu\nu}\longrightarrow h_{\mu\nu}+\xi_{\mu,\nu}+\xi_{\nu,\mu}$.)

An analogous procedure to establish energy-momentum conservation is also adopted in the ``Entwurf'' theory (see Sect. 4).

\subsubsection{The problem with the Newtonian limit}

The problem with the Newtonian limit was, it appears, one of the main reasons why Einstein abandoned the general covariance of the field equations. Apparently, (\ref{Eq:12}) did not reduce to the correct limit. That it leads to the Poisson equation for $g_{00}(\mathbf{x})$ is fine, but because of the harmonic coordinate condition the metric \textbf{can not be spatially flat}. (The almost Newtonian approximation of (\ref{Eq:10}) and (\ref{Eq:12}) is derived in textbooks on GR; see, e.g., \cite{Book}, Sect. 4.2.) Einstein found this unacceptable. He was convinced, I recall, that for (weak) static gravitational fields the metric must be of the form $(g_{\mu\nu})=\mbox{diag}(g_{00}(\mathbf{x}),1,1,1)$, as he already noted on p. 1 of his research notes. I wonder why he did not remember his cautious remark in one of his Prague papers \cite{Ein4} on static gravitational fields, in which -- while assuming spatial flatness -- he warned that this \textit{may very well turn out to be wrong}, and says that actually it does not hold on a rotating disk. Since a non-flatness would not affect the geodesic equation in the Newtonian limit, there is actually, as we all know, no problem. But Einstein realized this only three years later\footnote{In his calculation of the perihelion motion (on the basis of the vacuum equations $R_{\mu\nu}=0$) it became clear to him that spatial flatness did not hold even for weak static fields.}. Well(!): ``If wise men did not err, fools should despair'' (Wolfgang Goethe).

\section{The Einstein-Grossmann field equations}
\label{sec:3}

Einstein's difficulties, discussed previously, were among the reasons that he abandoned general covariance for the field equations. Another argument had to do with energy-momentum conservation. Generalizing the argument of Sect. 3.3.1 to the full theory, i.e., replacing $T^{\alpha\beta}$ in the second term of the conservation law (\ref{Eq:3}) 
for matter should lead to a conservation law for matter plus gravity of the form
\begin{equation}
\partial_{\nu}[\sqrt{-g}(T_{\mu}{}^{\nu}+t_{\mu}{}^{\nu})]=0.
\label{Eq:17}
\end{equation}
Now, Einstein thought at the time that the gravitational part  $t_{\mu}{}^{\nu}$ in a covariant theory should also be a tensor under general coordinate transformations.\footnote{This is my interpretation of a statement by Einstein in a lecture given to the Annual Meeting of the Swiss Naturforschende Gesellschaft on September 1913  (CPAE, Vol. 4, Doc. 16),  in which he says in connection of equation (\ref{Eq:17}) that the quantities $T_{\mu}{}^{\nu}$ and $t_{\mu}{}^{\nu}$ should have the same invariant-theoretical character. A similar statement is contained in a letter to Lorentz from August 16, 1913  (CPAE, Vol. 5, Doc. 470) and later in Sect. 6 of \cite{Ein5}.} This is, however, impossible, since then (\ref{Eq:17}) could not hold in all coordinate systems. (We know that Einstein obtained equation (\ref{Eq:17}) also for the final theory, but $t_{\mu}{}^{\nu}$ is then Einstein's \textit{pseudo-tensor}, as an expression of the equivalence principle. This caused, as is well-known, lots of discussions over decades. Energy-momentum conservation is really a delicate subject in GR, and has only a restricted meaning for isolated systems.)

Later, by August 1913, Einstein came up with yet another general argument, related to causality (`hole' argument). I shall discuss this in Sect. 7.

Having said this, I indicate now how Einstein arrived at the field equations of the `Entwurf' theory \cite{EG}. The starting point is again energy-momentum conservation, which we repeat
\[ \frac{1}{\sqrt{-g}}\partial_\nu(\sqrt{-g}g_{\mu\lambda}T^{\lambda\nu})-\frac{1}{2}\partial_{\mu}g_{\alpha\beta}T^{\alpha\beta}=0.\] 
The field equations are assumed to be of the form (\ref{Eq:4}), (\ref{Eq:5}), but now the covariance group is a priori not known. We only assume that it contains the general linear group. Therefore, covariance arguments do not much constrain the functional $\Gamma[g]$. Einstein hoped, of course, that some larger covariance group would emerge, which contains at least the non-linear transformations to uniformly accelerated frames.
 
Einstein makes an ansatz equivalent to (\ref{Eq:5}):
\begin{equation}
\Gamma^{\mu\nu}[g]=\frac{1}{\sqrt{-g}}\partial_{\alpha}(\sqrt{-g}g^{\alpha\beta}\partial_{\beta}g^{\mu\nu})+H^{\mu\nu}(g,\partial g),
\label{Eq:18}
\end{equation}
 because the first term would lead in linearized approximation to $\square h_{\mu\nu}$. The problem, discussed in Sect. 3.2, disappears if the coordinate condition $\partial^{\nu}h_{\mu\nu}=0$ is imposed. (Since the Ricci tensor is abandoned, the harmonic coordinate condition is no more needed.)
 
 Einstein replaces, as before, in the second term of the energy-momentum conservation law $T^{\alpha\beta}$ by $\Gamma^{\alpha\beta}[g]$, and tries to determine $H^{\mu\nu} (g,\partial g)$ such that  $\partial_{\mu}g_{\alpha\beta}\Gamma^{\alpha\beta}$ becomes a total divergence. He finds such an object by applying several partial integrations for the contribution of the first term in (\ref{Eq:18}). These manipulations are worked out on two pages of the research notes (pp. 51-52; see  \cite{Ein1}, pp. 262-263). The details are also presented by Grossmann in \cite{EG}, Sect. 4.3. Einstein gives the resulting identity in equation (12) of his part, and then writes down the corresponding ``Entwurf'' field equations and conservation laws. One of two equivalent forms given in the paper reads
\begin{equation}
\Gamma^{\mu\nu}[g]=\frac{1}{\sqrt{-g}}\partial_{\alpha}(\sqrt{-g}g^{\alpha\beta}\partial_{\beta}g^{\mu\nu})-g^{\alpha\beta}g_{\sigma\rho}\partial_{\alpha}g^{\mu\sigma}\partial_{\beta}g^{\nu\rho}-\kappa t^{\mu\nu},
\label{Eq:19}
\end{equation}
with 
\begin{equation}
-2\kappa t^{\mu\nu}=g^{\alpha\mu}g^{\beta\nu}\partial_{\alpha}g_{\sigma\rho}\partial_{\beta}g^{\sigma\rho}-\frac{1}{2}g^{\mu\nu}g^{\alpha\beta}\partial_{\alpha}g_{\sigma\rho}\partial_{\beta}g^{\sigma\rho}.
\label{Eq:20}
\end{equation}  
The conservation law (\ref{Eq:17}) holds with this expression for $t^{\mu\nu}$.\footnote{In GR $-\frac{1}{2}\partial_{\mu}g_{\alpha\beta}G^{\alpha\beta}=\frac{1}{\sqrt{-g}}\partial_{\nu}(\sqrt{-g}\kappa t_{\mu}{}^{\nu})$, where $ t_{\mu}{}^{\nu}$ is Einstein's pseudo-tensor. The resulting conservation law (\ref{Eq:17}) is compatible with Einstein's field equations, because the following identity holds: $\partial_{\nu}[\sqrt{-g}(G_{\mu}{}^{\nu}+\kappa t_{\mu}{}^{\nu})]=0$. This is actually equivalent to the contracted Bianchi identity $\nabla_{\nu}G_{\mu}{}^{\nu}=0$, which Einstein did not yet know at the time when he arrived at his final field equations. (We come back to this in Appendix A.)  In passing we recall that Einstein's pseudo-tensor is not symmetric. A useful symmetric pseudo-tensor was introduced by Landau and Lifshitz (see, e,g. \cite{Book}, Sect. 2.7). The use of pseudo-tensors has often be criticized, but at least mathematically these correspond to well-defined global geometrical objects on the frame bundle.}
 
Contrary to what Einstein and Grossmann claim in their joint paper, their procedure of constructing the field equations does \textit{not lead to a unique result}.
 
Einstein showed explicitly only later in 1913 in his famous Vienna lecture \cite{Ein5} that the Newtonian limit in his sense (with a flat spatial metric) is indeed recovered.
 
In collaboration with his lifelong friend Michele Besso, Einstein studied the perihelion motion of Mercury on the basis of the ``Entwurf'' theory. The result was 5/12 of what Einstein found later (1915)  for GR (CPAE, Vol. 4, Doc. 14).

\section{Further remarks on the two Einstein-Grossmann papers}
\label{sec:4}

\subsection{Interaction of matter with (external) gravitational fields}
\label{sec:4.1}

In Sect. 6 of \cite{EG} Einstein discusses the influence of (external) gravitational fields on matter. Beside the examples treated already earlier (geodesic equation of motion for point particles, energy-momentum balance of material systems), he generalizes Maxwell's equations to the generally covariant equation we all know. This part has survived in GR. The procedure is not yet formalized to the ``$\partial\longrightarrow\nabla $'' rule, as an expression of a local version of the equivalence principle. 

\subsection{Is a scalar theory of gravity possible?}
\label{sec:4.2}

In the final Sect. 7 of \cite{EG}, entitled as ``Can the gravitational field be reduced to a scalar?'', Einstein presents an ingenious Gedankenexperiment, which allegedly demonstrates that a Poincar\'{e}-invariant scalar theory of gravity, with a coupling of the scalar field to the trace of the energy-momentum tensor of matter, violates energy conservation. This can hardly be the case, because energy in such a theory has to be conserved, thanks to Noether's theorem. This general argument was not available to Einstein, since Noether's seminal paper appeared in 1918. For a careful critical analysis of Einstein's repeated reasoning against scalar gravity theories, proposed in particular by Nordstr\"om, I refer to \cite{NG}. Since Einstein's scalar theory, in collaboration with Fokker, will later be discussed, I leave it with that. 

\subsection{Variational principle and covariance group for the ``Ent\-wurf'' equations}
\label{sec:4.3}

In a second paper by Einstein and Grossmann \cite{EG2}, the authors investigate the covariance properties of their field equations, and show that the covariance group is larger than the linear group. As a tool they establish the following variational principle for their field equations:
\begin{equation}
\delta\int \mathcal{L}[g]\sqrt{-g}\,d^4x=\kappa\int T_{\mu\nu}\delta g^{\mu\nu}\sqrt{-g}\,d^4x\, ,
\label{Eq:20}
\end{equation}
with
\begin{equation}
\mathcal{L}[g]=-\frac{1}{2}g^{\alpha\beta}\partial_{\alpha}g_{\mu\nu}\partial_{\beta}g^{\mu\nu}.
\label{Eq:21}
\end{equation}

In later developments on the way to GR, variational principles were often used by Einstein, but -- before Hilbert -- he did not consider the curvature scalar.\footnote{Einstein gave the correct action for GR, that includes the boundary terms, in his 1916 paper ``Hamiltonian Principle and the General Theory of Relativity'' (CPAE, Vol. 6, Doc. 41). The surface terms are nowadays attributed to Gibbons-Hawking-Perry-York, etc. .}

\section{The Einstein-Fokker theory}
\label{sec:5}

During the Winter semester of 1913-1914, Adriaan D. Fokker, a student of Lorentz, visited Einstein in Z\"urich. The two collaborated on a non-linear
generalization of Nordstr\"om's theory, and came up with a consistent theory of gravity \cite{EF} that embodies the equivalence principle (actually the strong version, see \cite{Se}). Although it turned out that it is not viable empirically, the Einstein-Fokker theory is still interesting, mainly for pedagogical reasons. 

In a non-geometrical (flat-spacetime) formulation the Lagrangian is given by
\begin{equation}
\mathcal{L} = -\frac{1}{2} \partial_{\mu}\varphi\, \partial^{\mu}\varphi
         + \mathcal{ L}_{mat}\left[ \psi; (1 + k\varphi)^2\eta_{\mu\nu} \right] (1 + k\varphi)^4;
\end{equation}
in particular, the flat metric $\eta_{\mu\nu}$ in ${\cal L}_{mat}$ is replaced by
$(1 + k\varphi)^2 \eta_{\mu\nu}$, $k^2 = \kappa/2$.

One can replace the Minkowski metric by the ``physical metric'':
\begin{equation}
g_{\mu\nu} = (1 + k\varphi)^2 \eta_{\mu\nu}\, .
\end{equation}
For example, only relative to this metric the Compton wave
length is constant, i.e., not spacetime dependent.

Einstein and Fokker gave a geometrical formulation
of the theory. This can be summarized as follows:
\begin{enumerate}
\item[(i)]{ spacetime is conformally flat: Weyl tensor $=0$;}
\item[(ii)]{field equation: $R = 24\pi G\, T$;}
\item[(iii)]{ test particles follow geodesics.}
\end{enumerate}
In adapted coordinates, with $g_{\mu\nu} = \phi^2 \eta_{\mu\nu}$,
one finds
\begin{equation}
R = -6\phi^{-3}\eta^{\mu\nu} \partial_{\mu} \partial_{\nu}\phi,
\end{equation}
and the field equation becomes
\begin{equation}
\eta^{\mu\nu} \partial_{\mu}\partial_{\nu}\phi = -4\pi G \phi^3\, T.
\end{equation}

The Einstein-Fokker theory is generally covariant (as
emphasized in the original paper), however, {\it not} generally
{\it invariant}. I use this opportunity to point out the
crucial difference of the two concepts. For a long time
people (including Einstein) were not fully aware of this,
which caused lots of confusion and strange controversies.
(See, e.g., the preface of Fock's book on GR.)

The {\it invariance} group of a theory is the subgroup of the
covariance group that leaves the absolute, non-dynamical
elements of the theory invariant. (For a definition of this concept, see \cite{Book}, Sect. 2.5.)   In the Einstein-Fokker
theory the conformal structure is an {\it absolute} element: The object $\tilde g_{\mu\nu}=g_{\mu\nu}/(-g)^{1/4}$ is an absolute tensor density, in that it is diffeomorphic (as a tensor density) to $\eta_{\mu\nu}$. Therefore, the invariance group is the {\it conformal group},
which is a finite dimensional Lie group. In GR, on the other hand, the metric is entirely dynamical, and therefore the covariance group is, at the same time,
also the invariance group. In this sense, ``general relativity'' is an appropriate naming, never mind Fock and others.

Since the scalar theory of  Nordstr\"om and the generalization by Einstein and Fokker predict no
global light deflection\footnote{Einstein's equivalence principle implies that locally there is always light deflection, but as the Einstein-Fokker theory shows, this does not imply bending of light rays from a distant source traversing the gravitational field of a massive body and arriving at a distant observer.  (For a detailed discussion of how this comes about, see \cite{ER}.)}, Einstein urged in 1913 astronomers to measure the light deflection during the
solar eclipse in the coming year on the Crimea. Moreover, both predict -1/6 the Einsteinian value for the perihelion advance, in
contrast to observation.

For an extensive historical account of scalar gravitational theories, I refer to \cite{Nor3}.

\section{The `hole' argument against general covariance}
\label{sec:6}

At the time when he finished the paper with Grossmann, Einstein wrote to Ehrenfest on May 28, 1913:
``The conviction to which I have slowly struggled through is that \textit{there are no preferred coordinate systems of any kind}. However, I have only partially succeeded, even formally, in reaching this standpoint.'' (CPAE, Vol. 5, Doc. 441.) In a lecture given to the Annual Meeting of the Swiss Naturforschende Gesellschaft in September 1913, Einstein stated: ``It is possible to demonstrate by a general argument that equations that completely determine the gravitational field cannot be generally covariant with respect to arbitrary substitutions.'' (CPAE, Vol. 4, Doc. 16.)  He repeated this statement shortly afterwards in his Vienna lecture \cite{Ein5} of September 23, 1913.

The so-called ``hole'' (``Loch'') argument runs as follows (instead of coordinate transformations, I use a more modern language): Imagine a finite region $\mathcal{D}$ of spacetime -- the `hole' -- in which the stress-energy tensor vanishes. Assume that a metric field $g$ is a solution of generally covariant field equations. Apply now a diffeomorphism $\varphi$ on $g$, producing $\varphi_{*}g$ (push-forward), and choose the diffeomorphism such that it leaves the spacetime region outside $\mathcal{D}$ pointwise fixed. Clearly, $g$ and  $\varphi_{*}g$ are different solutions of the field equations that agree outside $\mathcal{D}$. In other words, generally covariant field equations allow huge families of solutions for one and the same matter distribution (outside the hole). At the time, Einstein found this unacceptable, because this was in his opinion a dramatic failure of what he called the law of causality (now usually called determinism). He then thought that the energy-momentum tensor should (for appropriate boundary or initial conditions) determine the metric \textit{uniquely}.

It took a long time until Einstein understood that this non-uniqueness is an expression of what we now call \textit{gauge invariance}, analogous to the local invariance of our gauge theories in elementary particle physics. On January 3, 1916 he wrote to Besso: ``Everything in the hole argument was correct up to the final conclusion''.

The role of diffeomorphism invariance of GR, especially for the Cauchy problem, was first understood by Hilbert. Being here at a cosmology conference, I do not have to explain that gauge invariance and gauge conditions play an everyday role in our theoretical studies, for example in cosmological perturbation theory (see, e.g., \cite{NS}).

\section{Concluding remarks}
\label{sec:7}

I break off this historical sketch at a point about two years before ``the final emergence into the light''. In an appendix I give a streamlined version of the arguments that led Einstein to modify the field equations (\ref{Eq:9}), which he reconsidered in November 1915, by the famous\textit{ trace term}.  

When Einstein was finishing his work on GR under great stress and was suspending all correspondence with colleagues, he still found time to communicate with Michele Besso. On November 17, 1915 he mailed a postcard from Berlin, that contains the great news:
\begin{quote}
 I have worked with great success during these months. \textit{General covariant} gravitational equations. \textit{Motions of the perihelion quantitatively explained}. Role of gravitation in the structure of matter [im Bau der Materie]. You will be amazed. I worked horribly strenuously [schauderhaft angestrengt], [it is] strange that one can endure that. (...) (CPAE, Vol. 8, Part A, Doc. 147).
\end{quote}
Besso passed this card on to Zangger: ``I enclose the historical card of Einstein, reporting the setting of the capstone of an epoch that began with Newton's `apple'.''

In a particularly instructive detailed technical letter of November 28, 1915 to Arnold Sommerfeld, Einstein summarizes his final struggle. Here just two crucial sentences from this important document \cite{Ein6}:
\begin{quote}
I realized ... that my previous gravitational equations were completely untenable. (...) After all confidence thus had been lost in the results and methods of the earlier theory, I saw clearly that only through a connection with the general theory of covariants, i.e., with Riemann's covariant [tensor], could a satisfactory solution be found. (...)
\end{quote}

The discovery of the general theory of relativity has often been justly praised as one of the greatest intellectual achievements of a human being. At the ceremonial presentation of Hubacher's bust of A.~Einstein in Z\"urich,  W.~Pauli said:
\begin{quote} 
The general theory of relativity then completed and - in contrast to the special theory - worked out by Einstein alone without simultaneous contributions by other researchers, will forever remain the classic example of a theory of perfect beauty in its mathematical structure.
\end{quote}

Let me also quote M.~Born:
\begin{quote} 
[The general theory of relativity] seemed and still seems to me at present to be the greatest accomplishment of human thought about nature; it is a most remarkable combination of philosophical depth, physical intuition and mathematical ingenuity. I admire it as a work of art.
\end{quote}

%%%%%%%%%%%%%%%%%%%%%%%%%%%%%%%%%%%%%%%%%%%%%%%%%%%%%%%%%%%%%%%%%%%%%%%%%%%%%%%%%%%%%%%%
\newpage

\begin{appendix}

\section{The final phase in November 1915}

In connection with the Einstein-Hilbert relation there have been lots of discussions and controversies on the famous trace term in the final field equations of GR. It may be of some interest if I present a streamlined version of Einstein's reasoning in November 1915. In what follows, I will rewrite Einstein's arguments, presented in CPAP, Vol. 6, Docs. 21, 22 and 25, without changing the content.\footnote{Since Ehrenfest had some technical difficulties in his study of these papers, Einstein presented in a letter all the calculational details \cite{Ein8}.} (Recall that he did not yet know the contracted Bianchi identity.)

We make use of the following identity between the Einstein tensor $G_{\mu}{}^{\nu}$ and Einstein's pseudo-tensor\footnote{We recall Einstein's definition of this pseudo-tensor \[2\kappa t_{\mu}{}^{\nu} := \delta_{\mu}{}^{\nu}L-g^{\alpha\beta}{}_{,\mu} \frac{\partial L} {\partial g^{\alpha\beta}{}_{,\nu}}, \] where $L= g^{\mu\nu}\Gamma^{\alpha}{}_{\mu\beta}\Gamma^{\beta}{}_{\nu\alpha}.$ In unimodular coordinates $L=R+g^{\alpha\beta}{}_{,\alpha\beta}.$}  $t_{\mu}{}^{\nu}$ in unimodular coordinates (always used in this note):
\begin{equation}
\underbrace{G_{\mu}{}^{\alpha}{}_{,\alpha}}_{R_{\mu}{} ^{\alpha}{}_{,\alpha} -(1/2)R_{,\mu}}+\;\kappa\, t_{\mu} {} ^{\alpha}{}_{,\alpha}=0\,.
\label{Eq:Nov1}
\end{equation}
This identity is implicitly contained in the cited papers\footnote{Here, the detailed justification of this claim. Einstein's equation above his formula (22) in Doc. 21 agrees with the identity 
\[ R_{\mu}{}^{\nu}=(g^{\nu\beta}\Gamma^{\alpha}{}_{\mu\beta})_{,\alpha}-\kappa (t_{\mu}{}^{\nu} -\frac{1}{2}\delta_{\mu}{}^{\nu} t)\]
($t=t_{\mu}{}^{\mu}$), if his proposed field equation is not used (i.e., if $\kappa T_{\mu}{}^{\nu}$ is replaces by $R_{\mu}{}^{\nu}$). Taking the divergence, the same calculation that led Einstein to his equation (22) gives the identity (\ref{Eq:Nov1}) above. Indeed, replacing $\kappa( T_{\mu}{}^{\nu}+ t_{\mu}{}^{\nu})_{,\nu}$ by 
$( R_{\mu}{}^{\nu}+\kappa t_{\mu}{}^{\nu})_{,\nu}$, and noting that his left-hand side is just $R_{,\mu}$, one obtains identity (\ref{Eq:Nov1}). In other words, Einstein's derivation of $R_{,\mu}=0$ on the basis of his proposed field equation in Doc. 21 can be interpreted as a derivation of (\ref{Eq:Nov1}).}, and later derived in Einstein's first review paper Doc. 30.

In Doc. 21 (submitted on 4 November, 1915) and the Addendum Doc. 22 (11 November), Einstein came back to the field equation $R_{\mu\nu}=\kappa T_{\mu\nu}$, he had already considered with Grossmann three years before. Basically, he argues as follows: 

1. The identity (\ref{Eq:Nov1}), together with the field equation imply
\begin{equation}
\kappa(T_{\mu}{}^{\alpha}+t_{\mu}{}^{\alpha})_{,\alpha}=\frac{1}{2}R_{,\mu}.
\label{Eq:Nov2}
\end{equation}

2. Einstein derives the identity
\begin{equation}
\frac{1}{2} g^{\mu\nu}{}_{,\lambda}R_{\mu\nu} = \kappa t_\lambda{}^{\nu}{}_{,\nu} ,
\label{Eq:Nov3}
\end{equation}
uses on the left the field equation, and then the energy-momentum law $\frac{1}{2} g^{\mu\nu}{}_{,\lambda}T_{\mu\nu}=- T_\lambda{}^{\nu}{}_{,\nu}$, to conclude that
\begin{equation}
( T_\mu{}^{\nu} + t_\mu{}^{\nu} )_{,\nu}=0\,.
\label{Eq:Nov4}
\end{equation}
Together, one obtains $R_{,\mu}=0$, hence from the trace of the field equation $T$=const.

At first, Einstein regarded this as an interesting restriction on a theory of matter. We recall that in those years an electromagnetic world view, based on a nonlinear generalization of Maxwell's theory, was quite popular. (An unsuccessful, but much discussed example, had been put forward in 1912 by Gustav Mie.) But soon, Einstein wanted to get rid of this restriction.

Looking back at the arguments above, it is quite obvious how to achieve this. Thanks to the identity (\ref{Eq:Nov1}), we obtain in the first step also the conservation law (\ref{Eq:Nov4}), if the field equation 
\begin{equation}
  G_{\mu\nu}=\kappa T_{\mu\nu}
\end{equation}
is adopted. The second step implies this too, because for unimodular coordinates the left-hand side of (\ref{Eq:Nov3}) is unchanged if $R_{\mu\nu}$ is replaced by $G_{\mu\nu}$. So, no additional constraint is imposed. Moreover, using the result from step 1, we conclude in step 2 that the energy-momentum law $T_{\mu}{}^{\nu}{}_{;\nu}=0$ \textit{follows from the field equation}. This is, of course, stressed by Einstein, and also that no additional constraints on any matter theory are imposed by gravity.

In view of this approach to the field equation, I see no reason to suspect that Einstein took the trace term from Hilbert.

Einstein was well aware that his original presentation was not yet satisfactory. To Lorentz he wrote on January 17, 1916 \cite{Ein7} : "The basic equations are now finally good, but the derivations abominable; this drawback still has to be removed."

\section{On Einstein's first review paper on GR}

In what follows, I also rewrite Einstein's arguments, presented in CPAP, Vol. 6, Doc. 30, again without changing the content. (Recall that he still did not know the contracted Bianchi identity.)

We begin with a well-known identity between the Einstein tensor $G_{\mu}{}^{\nu}$ and Einstein's pseudo-tensor  $t_{\mu}{}^{\nu}$ in unimodular coordinates (always used in this note):
\begin{equation}
G_{\mu}{} ^{\alpha}+\kappa\, t_{\mu} {} ^{\alpha}=\frac{1}{2} U_{\mu} {} ^{\alpha\beta} {} _{,\beta}\,,
\label{Eq:A1}
\end{equation}
with the super-potential (Freud)
\begin{equation}
U_{\mu} {} ^{\alpha\beta} = g_{\mu\sigma}H^{\sigma\rho\alpha\beta}{}_{, \rho}\,, \quad H^{\sigma\rho\alpha\beta}=g^{\sigma\alpha}g^{\rho\beta}- g^{\sigma\beta}g^{\rho\alpha}\,.
\end{equation}
(It is not complicated to derive this identity with the tools developed in Sect. 15 of the cited document.)

From this it follows that the vacuum equation $R_{\mu\nu}=0$ can be written as 
\begin{equation}
\frac{1}{2} U_{\mu} {} ^{\alpha\beta} {} _{,\beta}=\kappa\, t_{\mu} {} ^{\alpha}\,.
\end{equation}
This is equivalent to what Einstein does in a first step. The obvious identity $U_{\mu} {} ^{\alpha\beta} {} _{,\beta\alpha}\equiv 0$ implies that $t_{\mu}{}^{\alpha}{}_{,\alpha}=0$, whence $t_{\mu}{}^{\alpha}$ is interpreted by Einstein as the energy-momentum complex (pseudo-tensor) of the gravitational field.

In the presence of matter, Einstein just replaces $t_{\mu}{}^{\alpha}$ by the sum $ t_{\mu}{}^{\alpha} +T_{\mu}{}^{\alpha}$, obtaining the field equations
\begin{equation}
\frac{1}{2}U_{\mu} {} ^{\alpha\beta} {} _{,\beta}=\kappa\, ( t_{\mu} {} ^{\alpha} + T_{\mu}{}^{\alpha})\,.
\end{equation}
These guarantee the conservation law $( t_{\mu} {} ^{\alpha} + T_{\mu}{}^{\alpha})_{,\alpha}=0$. By the identity above this form is equivalent to 
$G_{\mu}{} ^{\alpha}=\kappa\, T_{\mu} {} ^{\alpha}$, with the correct trace term. We also note that the identity $U_{\mu} {} ^{\alpha\beta} {} _{,\beta\alpha}\equiv 0$ is equivalent to the contracted Bianchi identity. To see this, one must also use the relation (\ref{Eq:Nov3}).

His successful explanation of Mercury's perihelion precession had convinced Einstein that his gravitational vacuum equations were definite. It is, therefore, natural that he began his reasoning with a physical interpretation of this solid basis.

\end{appendix}

\section*{Acknowledgements}
I thank the organizers of the stimulating conference ``Cosmology since Einstein'' for inviting me to discuss Einstein's great progress in his search for GR after his second return to Z\"urich in summer 1912. I am very grateful to Domenico Giulini for detailed constructive criticism of an earlier version of the manuscript, and clarifying discussions. Thanks go to Friedrich Hehl and G\"unther Rasche for a careful reading of the manuscript.


\begin{thebibliography}{}
%
\bibitem{Ein0}
CPAE,  \emph{The collected papers of Albert Einstein}, Edited by J. Stachel {\it et al.}, Vols. 1-12.
Princeton: Princeton University Press, 1987--2010.
%
\bibitem{EqP}
A. Einstein, On the Relativity Principle and the Conclusions Drawn from It, CPAE, Vol. 2, Doc. 47.
%
\bibitem{GS}
D. Giulini and N. Straumann, Einstein's Impact on the Physics of the Twentieth Century, \textit{Studies in the History and Philosophy of Modern Physics}, 37: 115-173, 2006 [arXiv: physics/0507107].
%
\bibitem{Ein3}
A. Einstein, Erinnerungen-Souveniers, \textit{Schweizerische Hochschulzeitung 28 (Sonderheft)} (1955), pp. 145-153; reprinted as ``Autobiographische Skizze,'' in Carl Seelig, ed., \textit{Helle Zeit-Dunkle Zeit. In Memoriam Albert Einstein, Z\"urich, Europa Verlag, 1955}, pp. 9-17.
%
\bibitem{EqP2}
A. Einstein, On the Influence of Gravitation on the Propagation of Light, CPAE, Vol. 3, Doc. 23.
%
\bibitem{Ein4a}
A. Einstein, On the Theory of the Static Gravitational Field, CPAE, Vol. 4, Doc. 4.
%
\bibitem{Ein1}
A. Einstein, Research Notes on a Generalized Theory of Gravitation, August 1912, CPAE, Vol. 4, Doc. 10.
%
\bibitem{Ein2}
A. Einstein, Letter to H.A. Lorentz, CPAE, Vol. 5, Doc. 470.
%
\bibitem{Nor1}
J. Norton, How Einstein Found his Field Equations: 1912-1915. In \textit{Historical Studies in the Physical Sciences}, J.L. Heilbron, ed., Vol. 14, pp. 253-316. Berkeley, University of California Press (1984).
%
\bibitem{Renn}
J. Renn, ed.,  \textit{General Relativity in the Making: Einstein's Z\"urich  Notebook}, Commentary and Essays. (\textit{The Genesis of General Relativity}, Vol. 2.) Dortrecht: Springer (2006).
%
\bibitem{Koll}
L. Kollros, Erinnerungen-Souveniers, \textit{Schweizerische Hochschulzeitung 28 (Sonderheft)} (1955), pp. 169-173; reprinted as ``Erinnerungen eines Kommilitonen,'' in Carl Seelig, ed., \textit{Helle Zeit-Dunkle Zeit. In Memoriam Albert Einstein, Z\"urich , Europa Verlag, 1955}, pp. 17-31.
%
\bibitem{EG}
A. Einstein and M. Grossmann, Outline of a Generalized Theory of Relativity and a Theory of Gravitation, CPAE, Vol. 4, Doc. 13. 
Entwurf einer verallgemeinerten Relativist\"atstheorie und einer Theorie der
Gravitation, \emph{Zeitschrift f\"ur Mathematik und Physik}, \textbf{62}, 225-259 (1914).  
%
\bibitem{NS}
N. Straumann, From Primordial Quantum Fluctuations to the
Anisotropies of the Cosmic Microwave Background Radiation, \textit{Ann.
Phys.} (Leipzig) \textbf{15}, No. 10-11, 701-847 (2006);
[arXiv: hep-ph/0505249]. For an updated and expanded version,
see: www.vertigocenter.ch/straumann/norbert
% 
\bibitem{Nor2}
J. Norton, `Nature is the Realization of the Simplest Conceivable Mathematical Ideas': Einstein and the Canon of Mathematical Simplicity, \textit{Stud. Hist. Phil. Mod. Phys.}, Vol. 31, pp. 135-170 (2000).
%
\bibitem{Book}
N. Straumann, General Relativity, With Applications to
Astrophysics, \textit{Texts and Monographs in Physics}, Springer Berlin, 2004.
%
\bibitem {Ein4}
A. Einstein, The Speed of Light and the Statics of the Gravitational Field, CPAE, Vol. 4, Doc. 3.
%
\bibitem{Ein5}
A. Einstein, On the Present State of the Problem of Gravitation, CPAE, Vol. 4, Doc. 17.
%
\bibitem{NG}
D. Giulini, What is (not) Wrong with Scalar Gravity?, \textit{Studies in the History and Philosophy of Modern Physics}, Vol. 39, pp. 135-180 (2008).
%
\bibitem{EG2}
A. Einstein and M. Grossmann, Covariance Properties of the Field Equations of the Theory of Gravitation Based on the Generalized Theory of Relativity, CPAE, Vol. 6, Doc. 2.
%
\bibitem{EF}
A. Einstein and A.D. Fokker, Nordstr\"om's Theory of Gravitation from the Point of the Absolute Differential Calculus, CPAE, Vol. 4, Doc. 28. 
Die Nordstr\"omsche Gravitationstheorie vom Standpunkt des Absoluten
Differentialkalk\"uls, \emph{Ann. Phys. (Leipzig)} \textbf{44}, 321-328 (1914).
%
\bibitem{Se}
R.U. Sexl, Theories of Gravitation, Fortschritte der Physik \textbf{15}, 269-307 (1967).
%
\bibitem{ER} 
J.~Ehlers and W.~Rindler, \textit{Gen. Rel. Grav.} \textbf{29}, 519 (1997).
%
\bibitem{Nor3}
J. Norton,  Einstein, Nordstr\"om and the Early Demise of
Scalar, Lorentz-Covariant Theories of Gravitation. \emph{Archive for
the History of Exact Sciences} \textbf{45}, 17-97 (1992).
%
\bibitem{Ein6}
Letter of Einstein to Arnold Sommerfeld, CPAE, Vol. 8, Part A, Doc. 153.
%
\bibitem{Ein7}
Letter of Einstein to H.A. Lorentz, CPAE, Vol. 8, Part A, Doc. 183.
%
\bibitem{Ein8}
Letter of Einstein to P. Ehrenfest, CPAE, Vol. 8, Part A, Doc. 185.

\end{thebibliography}
\end{document}